\documentclass[aps,prl,twocolumn,superscriptaddress]{revtex4}

\usepackage[dvips]{graphicx}

\newcommand{\HoYLF}{\text{LiHo}_x\text{Y}_{1-x}\text{F}_4}
\newcommand{\ham}{\mathcal{H}}

\begin{document}

\title{Specific Heat of the Dilute Ising Magnet $\HoYLF$}

\author{J. A. Quilliam}
\author{C. G. A. Mugford}

\affiliation{Department of Physics and Astronomy, University of
Waterloo, Waterloo, ON N2L 3G1 Canada}

\affiliation{Institute for Quantum Computing, University of
Waterloo, Waterloo, ON N2L 3G1 Canada}

\author{A. Gomez}
\author{S. W. Kycia}
\affiliation{Deparment of Physics, University of Guelph, Guelph, ON
N1G 2W1 Canada}

\author{J. B. Kycia}

\affiliation{Department of Physics and Astronomy, University of
Waterloo, Waterloo, ON N2L 3G1 Canada}

\affiliation{Institute for Quantum Computing, University of
Waterloo, Waterloo, ON N2L 3G1 Canada}

\date{\today}

\begin{abstract}

We present specific heat data on three samples of the dilute Ising
magnet $\HoYLF$ with $x = 0.018$, 0.045 and 0.080. Previous
measurements of the ac susceptibility of an $x = 0.045$ sample
showed the Ho$^{3+}$ moments to remain dynamic down to very low
temperatures~\cite{Reich1990,Ghosh2002} and the specific heat was
found to have unusually sharp features~\cite{Reich1990,Ghosh2003}.
In contrast, our measurements do not exhibit these sharp features in
the specific heat and instead show a broad feature, for all three
samples studied, which is qualitatively consistent with a spin glass
state. Integrating $C/T$, however, reveals an increase in residual
entropy with lower Ho concentration, consistent with recent Monte
Carlo simulations showing a lack of spin glass transition for low
$x$~\cite{Snider2005}.

\end{abstract}


\maketitle


Extensive work has previously been done to understand the spin glass
transition found in disordered magnetic systems~\cite{Binder1986}
and changes in behavior as the concentration of magnetic moments is
reduced~\cite{ListofSGPapers}.  The material $\HoYLF$ is a nearly
perfect example of a dilute, dipolar-coupled Ising magnet and is
therefore an ideal system for experimentally testing theories of
simple, interacting spin models. Despite the apparent simplicity of
this system's underlying model, however, a series of surprising
results and fascinating effects has emerged from the material's rich
phase diagram, especially at low concentrations of magnetic
Ho$^{3+}$ ions~\cite{Reich1990,45percentPapers,Ghosh2002,Ghosh2003}.

At $x=1$ the system has been found to order ferromagnetically with
$T_c = 1.53$~K~\cite{Bitko1996}, but below a certain amount of
dilution ($x\simeq 0.25$), there is enough randomness and
frustration (due to the angle-dependent dipolar interaction) that
the system becomes a spin glass~\cite{16percentPapers,Reich1990}. At
$x=0.045$, however, ac susceptibility experiments have shown the
material not to freeze down to very low
temperatures~\cite{Reich1990}.  The absorption spectrum
$\chi''(\omega)$ is observed to narrow with lower temperature where
typically, in a spin glass, the absorption spectrum becomes wider as
freezing of the moments leads to longer relaxation
times~\cite{Hueser1983}.  Furthermore, at temperatures below 100 mK,
there appears to be a gap in the absorption spectrum and coherent,
low frequency oscillations with lifetimes of up to 10~s are
observed.  These effects have been attributed to clusters of roughly
260 Ho ions acting as largely independent
oscillators~\cite{Ghosh2002}.

\begin{figure}
\begin{center}
\includegraphics[width=3.325in,keepaspectratio=true]{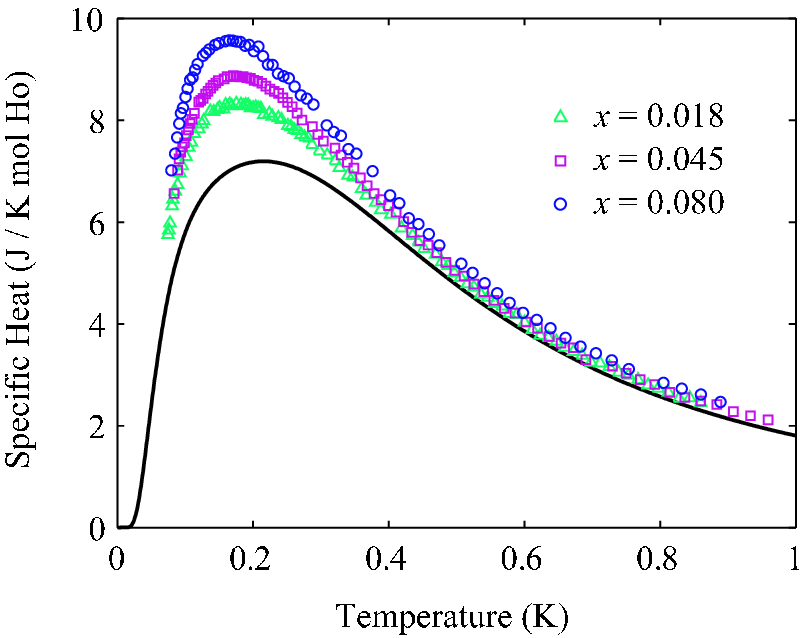}

\caption{\label{TotalSpecificHeat} Total measured specific heat for
$x=0.018$, $x=0.045$ and $x=0.080$.  The solid line is the total
non-interacting specific heat calculated by diagonalizing the
crystal field and nuclear hyperfine Hamiltonians.}

\end{center}
\end{figure}

Since the dipolar coupling between the Ho moments is a long-range
interaction, it has long been theoretically expected that there
should be no finite concentration of moments ($x$) at which the
ordering (or freezing) temperature of the system drops to
zero~\cite{Stephen1981}.  The ac susceptibility experiments
performed on the $x=0.045$ sample, however, seem to contradict this
theory as there is no sign of freezing even down to 50 mK.  Recent
Monte Carlo simulations of dipolar-coupled Ising moments randomly
placed on a cubic lattice do suggest that there is no spin glass
transition for $x<x_C \simeq 0.20$ which could explain the unusual
dynamics seen at $x=0.045$~\cite{Snider2005}.  There is also
evidence in recent $\mu$SR experiments on
LiHo$_{0.045}$Y$_{0.955}$F$_4$ of spin dynamics persisting down to
low temperatures~\cite{Rodriguez2006}.

This unique spin liquid or ``anti-glass'' state has also exhibited
unusually sharp features in its specific heat at around 110 mK and
300 mK.  These features were qualitatively reproduced in numerical
simulations using a model based on quantum entanglement of pairs of
moments and a pair-wise `decimation' procedure in which the sharp
heat capacity features correspond to maxima in the distribution of
dipolar couplings in the system~\cite{Ghosh2003}.  This simulation
was also able to reproduce a $T^{-0.75}$ behavior of the dc
susceptibility which was observed experimentally. It is not known
whether there is a relation between these heat capacity
signatures and the anomalous dynamics observed by ac susceptibility.


In this letter we present specific heat data taken on three
stoichiometries in this series: $x=0.018$, $0.045$ and $0.080$. The
samples studied are high quality single crystals grown with the
Bridgman technique~\cite{TydexFootnote}. Crystalline quality was
verified by high resolution diffraction on a fine focus Cu rotating
anode generator equipped with a high resolution Ge (220)
four-crystal monochromator and a Huber 4-circle diffractometer. The
measurements revealed extremely sharp Bragg peaks
($\theta_\mathrm{FWHM}<0.015^\circ$) for all reflections, indicating
high crystalline perfection. No twinning was observed. Extensive
diffuse scattering measurements revealed no diffuse scattering near
or away from the Bragg peaks, or satellite peaks that could be
associated with any disorder or short range ordering. Small $\sim
100$ $\mu$m fragments were taken from each sample and
crystallography data sets were measured using a molybdenum rotating
anode, kappa diffractometer and CCD area detector. All three data
sets refined well with Ho substituting for Y in the expected
tetragonal ($I4_1/a$) structure~\cite{Garcia1993}.

Measurements were performed using the quasi-adiabatic method with a
long time-constant $\tau$ of relaxation.  No substrate was used in
these experiments and the heater, thermometer and weak link were
glued directly onto the sample which was suspended from very fine
nylon threads.  Using a substrate can lead to an underestimate of
the heat capacity due to large thermal resistances between the
sample and substrate.  The addendum was determined to be less than
0.1\% of the sample's heat capacity. Samples were typically discs
$\sim 8$~mm in diameter and $\sim 1$~mm thick.

A RuO$_2$ resistor (1 k$\Omega$ at 300 K) was used as a thermometer
and a 10 k$\Omega$ metal-film resistor was used as a heater, both
with thinned alumina substrates.  Leads to the thermometer and
heater were 6 $\mu$m diameter, $\sim 5$ mm, NbTi, superconducting
wires with a thermal conductance of $K_\mathrm{NbTi} \simeq 8\times
10^{-11}$ W/K at 1 K and at least a factor of 10 smaller at 100 mK.
The thermal conductance from the thermometer and heater to the
sample ($K_\mathrm{TS}$ and $K_\mathrm{HS}$ respectively) were
measured to be greater than $10^{-8}$ W/K at very low $T$ ($<50$
mK). The weak link connecting the sample to the dilution
refrigerator mixing chamber was made from manganin wire and had a
thermal conductance $K_\mathrm{WL} \simeq 1\times 10^{-7}$ W/K at
100 mK. Calculations show that the temperature of the thermometer
differs from that of the sample by less than 0.1\%.

\begin{figure}
\begin{center}
\includegraphics[width=3.325in,keepaspectratio=true]{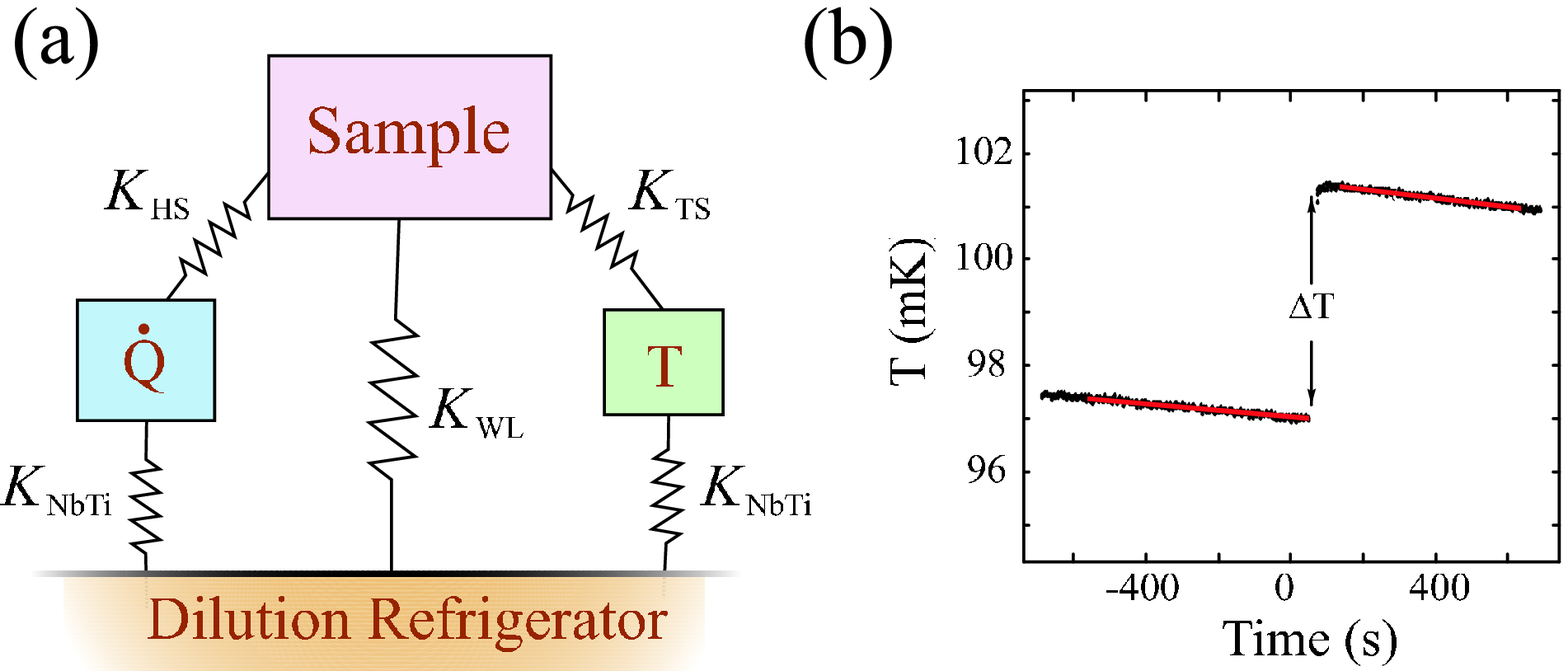}
\caption{\label{ExperimentFigure} Diagram of relevant thermal links
in experimental apparatus (a) and an example heat pulse showing
linear fits and extrapolation to the mid-point of the pulse (b).}
\end{center}
\end{figure}

Thermometer resistance measurements were made with a LR-700 AC
Resistance Bridge.  The cell was contained in a copper radiation
shield and the cryostat was surrounded by a lead shield and two
$\mu$-metal shields to attenuate the external magnetic field. The
thermometer resistance was consistent with a standard RuO$_2$
temperature dependence~\cite{Pobell1992} with no indication of
self-heating in the range of our data. The RuO$_2$ thermometer was
calibrated to a calibrated \emph{LakeShore} Ge resistance
thermometer and a CMN susceptibility thermometer.

Time constants on the order of several hours (at the lower
temperatures) ensured that the sample was cooled very slowly and was
therefore able to reach equilibrium.  Cooling the sample even more
slowly did not have a noticeable effect on the measured heat
capacity.  Temperature data was collected for up to 30 minutes on
either side of the heat pulse and the heat capacity is given by $C =
\dot{Q}/\Delta T$ where $\Delta T$ is obtained through extrapolation
to the midpoint of the pulse as is shown in
Figure~\ref{ExperimentFigure}(b).


For temperatures below 1 K, the specific heat of $\HoYLF$ is
dominated by a broad feature which arises from the $I=7/2$
nuclear-spin degrees of freedom.  The single-ion Hamiltonian
(neglecting the dipolar and nearest-neighbor exchange interactions)
is given by $\ham = \ham_\mathrm{CF} + \ham_\mathrm{HF} +
\ham_\mathrm{Q}$.  The $4f$-electrons in Ho$^{3+}$ are tightly
bound, resulting in a significant nuclear-hyperfine interaction:
$\ham_\mathrm{HF} = A\mathbf{I}\cdot\mathbf{J}$.  $\ham_\mathrm{Q}$
is the nuclear quadrupole interaction and $\ham_\mathrm{CF} =
\sum_{l,m} B_l^m O_l^m$ is the crystal field potential caused by
surrounding ions (the $O_l^m$'s are Steven's operator equivalents).
\begin{figure}
\begin{center}
\includegraphics[width=3.325in,keepaspectratio=true]{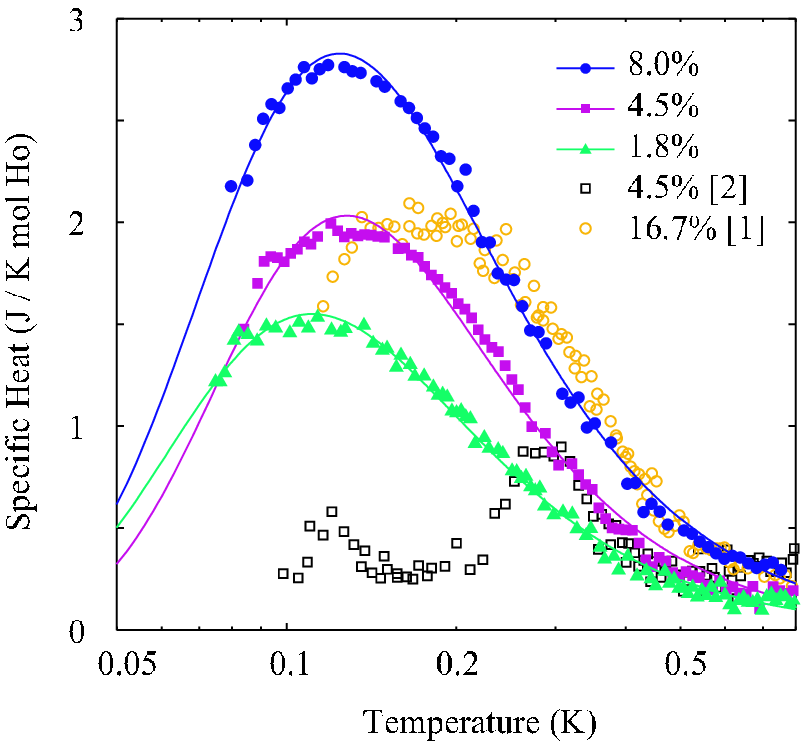}
\caption{\label{SubstractedDataFigure} Electronic moments'
contribution to the specific heat (solid line from
Figure~\ref{TotalSpecificHeat} subtracted) for $x=0.018$, $x=0.045$
and $x=0.08$ from this work (filled symbols).  Also $x=0.045$ from
Ghosh \emph{et al.}~\cite{Ghosh2003} and $x=0.167$ from Reich \emph{et al.}~\cite{Reich1990} (open
symbols).  The solid lines are fits of the form of Equation~\ref{Schottky}.}
\end{center}
\end{figure}

If the crystal field Hamiltonian is diagonalized by itself, one
obtains a ground-state, Ising doublet with an effective $g$-factor
of 13.8 and a next excited state at around 11~K.  It is then, in
some cases, safe to assume that the electronic moments are perfect
Ising spins and the specific heat can be expressed as the sum of an
electronic contribution $\Delta C$ and a nuclear contribution
\begin{equation}\label{CHF}
\frac{C_\mathrm{Nuclear}}{R} = \left( \frac{\sum_m x_m
e^{-x_m}}{\sum_m e^{-x_m} } \right)^2 - \frac{\sum_m x_m^2
e^{-x_m}}{\sum_m e^{-x_m} }, \end{equation}
where $x_m = -A_Jg_\mathrm{eff}m/2 g_J T + Pm^2/T$ and $m=-7/2,\ldots,7/2$.
A fit of this form was successfully applied by Mennenga {\it et
al.} to the specific heat of LiHoF$_4$ below the transition
temperature~\cite{Mennenga1984}. We have made corrections to this form by
diagonalizing the entire non-interacting Hamiltonian
(a $136\times 136$ matrix).  We have used
$A_J/k_B = 40.21$ mK, determined by EPR experiments on
LiHo$_{0.02}$Y$_{0.98}$F$_4$~\cite{Magarino1976}.  This is similar to the value $A_J/k_B =
39.8$ mK found for the pure material~\cite{Magarino1980}.  We have
assumed an axially symmetric nuclear quadrupole interaction of
strength $P=1.7$ mK which was determined with EPR on free Ho$^{3+}$
ions~\cite{Abragam1970}.  The crystal field parameters $B_l^m$ were
taken from~\cite{Chakraborty2004}.  The resulting single-ion
specific heat is shown as the solid line in
Fig.~\ref{TotalSpecificHeat} and has been subtracted from the data
to give $\Delta C$ in Fig.~\ref{SubstractedDataFigure}.  This more
detailed calculation of the non-interacting specific heat is lower
than Equation~\ref{CHF} by $\sim 4$\% near the highest point of the
curve.

As mentioned earlier, the validity of this subtraction depends on
the assumption that the Ho$^{3+}$ ions are perfect Ising moments.
This is not entirely the case, as the nuclear hyperfine interaction
introduces mixing with the next excited states.  Indeed, the nuclear
hyperfine interaction has been shown to strongly effect the magnetic
ordering of the pure material in transverse
field~\cite{Chakraborty2004} and the diluted material in the spin
glass regime~\cite{Schechter2005}.  Nevertheless, the total specific
heat should approach this non-interacting specific heat at higher
temperatures (close to 1 K).  A small phonon contribution to the
specific heat ($\propto T^3$) is also subtracted, estimated from
the heat capacity of the pure material above
1~K~\cite{Mennenga1984}.

The crystal supplier provided us with nominally 2\%, 4.5\% and 8\%
concentrations of holmium.  Assuming the correct nuclear hyperfine
component, however, the 2\% sample appears to be closer to 1.8\% as
the remaining term $\Delta C$ should behave as $T^{-2}$ at higher
temperatures (close to 1 K).  In this way, the 4.5\% and 8\% samples were
confirmed to be the correct stoichiometry.

In all three samples, the remaining specific heat contribution
$\Delta C$ is qualitatively similar and found to be a broad feature
which is somewhat consistent with the heat capacity of a spin
glass~\cite{Wenger1976,Meschede1980}. The specific heat of a spin
glass is not expected to show a remarkable feature at the spin glass
freezing transition as the critical exponent $\alpha$ is often
negative, in the range -2 to -4~\cite{Fischer1991}. Instead of
probing the actual freezing transition, the spin glass heat capacity
is more indicative of excitations above the transition. The simplest
situation is one excited energy state $E_1$ above the ground state
having a degeneracy $n$ with respect to the ground state degeneracy.
We can then apply fits of the form
\begin{equation}\label{Schottky}
\Delta C = C_0\frac{n(E_1/k_BT)^2e^{-E_1/k_BT}}{(1 -
ne^{-E_1/k_BT})^2} \end{equation} to the data. The resulting fitting
parameters for this data and an $x=0.167$ sample measured by Reich
\emph{et al.}~\cite{Reich1990} are shown in Table~\ref{FitTable}.
Clearly the size of the specific heat features decreases with
decreasing concentration $x$.  The peak temperature of the curve,
however, is very close in all three samples and does not appear to
scale with the Curie temperature ($xT_C$).

\begin{table}
\caption{\label{FitTable} Fitting parameters for $\Delta C$ for
$x=0.018$, $x=0.045$ and $x=0.08$ from this work and data taken from
Reich \emph{et al.} ($x=0.167$)~\cite{Reich1990}.  The peak temperatures $T_\mathrm{peak}$, the relative
width of the specific heat curve $\mathrm{FWHM}/T_\mathrm{max}$ and the
measured residual entropy $S_0$ (assuming a linear temperature
dependence at low $T$) are also given.}

\begin{ruledtabular}

\begin{tabular}{lllll}
Parameter      & 1.8\% & 4.5\%  & 8.0\%    & 16.7\% \\
\hline

$E_1/k_B$ (K)  & 0.26       & 0.32    & 0.29       & 0.46 \\
$n$            & 0.85      & 1.43       & 0.86        & 1.89 \\

$T_\mathrm{peak}$ (K)  & 0.11  &  0.13  &  0.12 &  0.17 \\

$\mathrm{FWHM}/T_\mathrm{peak}$ & 1.7 & 1.6 & 1.7   & 1.5 \\

$S_0/R$          & $0.31$  & $0.21$   & $0.00$    & $0.18$ \\

\end{tabular}

\end{ruledtabular}
\end{table}

Numerically integrating $\Delta C/T$ with respect to $T$ gives the
total amount of entropy released over the temperature range of these
measurements.  The total high-temperature entropy of an Ising magnet
is $R\ln2$ but there may be a residual ground state entropy seen in
doing this integral.  Lower temperature data is required in order to
confidently observe all the release of entropy in the system.  We
have extrapolated the data to 0 K, assuming a linear temperature
dependence, before integrating $\Delta C/T$.  Measurements at lower
temperatures must be made in order to determine the temperature
behavior of the specific heat below the maximum, but many past
measurements have observed a linear temperature dependence in spin
glasses~\cite{Wenger1976,Meschede1980} as described by the two-level
system (TLS) argument~\cite{Anderson1972}.

For the 8\% sample, this integral reveals approximately all of the
total expected entropy.  In the 1.8\% and 4.5\% samples, however,
our measurement observes a smaller percentage of $R\ln2$: 56\% and
70\% respectively, leaving a significant residual entropy $S_0$. The
residual entropy for each sample is also shown in
Table~\ref{FitTable}. These values may be compared to previous
measurements for $x=0.167$ where 75\% of the entropy was measured
(also assuming a linear temperature dependence below the peak) and
$x=0.045$ where only 15\% of $R\ln2$ was observed over the range of
the measurement~\cite{Reich1990,Ghosh2003}.  In the case of the
4.5\% and 16.7\% samples, $S_0$ is quite close to the value of
$0.199R$ predicted by the SK model of a spin
glass~\cite{Tanaka1980}.

The Monte Carlo simulations of Snider \emph{et al.} on a dilute
dipolar-coupled Ising system predict 0 residual entropy at
$x=0.20$~\cite{Snider2005}.  For $x<0.20$ they see no spin glass
ordering and an increasing $S_0$ with decreasing $x$ as a larger
number of degenerate grounds states are available. This is the same
trend observed in our data, though experimentally, the magnitude of
the residual entropy is much larger.  In the real system, the point
at which spin glass ordering ceases must be lower than $x=0.167$ as
this has been observed to be a spin glass~\cite{Reich1990} and may
be closer to $x=0.080$ at which point we have observed the release
of nearly all the expected entropy.


The relative broadness of the observed features may be parametrized
by the full width at half maximum (FWHM) divided by the peak
temperature $T_\mathrm{max}$.  The three samples studied here give
values around 1.7 (see Table~\ref{FitTable}).  This parameter is
approximately 1.2 in Au\emph{Fe}~\cite{Martin1980b} and 1.5 in
Eu$_x$Sr$_{1-x}$S for example~\cite{Meschede1980}.  Typically, the
maximum in the specific heat of spin glasses is found to be
approximately 20\% higher than the spin glass transition temperature
which is determined by ac susceptibility
experiments~\cite{Binder1986}.  If this rule of thumb were to apply
here, it would give spin glass transition temperatures of 90 to 100
mK for these three samples.

We have measured the specific heat of three samples at and around a
concentration of 4.5\% holmium, and our measurements do not exhibit
the sharp features that were seen
previously~\cite{Reich1990,Ghosh2003}.  The data sets agree well
until $\sim 300$ mK which indicates that there is no error in
stoichiometry.  Below this point, however, there is a significant
discrepancy.  Our data, therefore, also does not support the
theoretical model presented by Ghosh \emph{et al.}~\cite{Ghosh2003}.
We have taken great care to rule out any experimental errors such as
decoupling of the thermometer from the sample.  Recent thermal
conductivity measurements of LiHo$_{0.04}$Y$_{0.96}$F$_4$ also do
not show any remarkable features at 110 mK and 300
mK~\cite{Nikkel2005}.  In some systems which have $\chi\sim
T^{-\alpha}$ the specific heat also shows a simple power law
behavior with a related exponent~\cite{Brunner1997}. It would be
interesting to measure $C(T)$ to lower temperatures to look for such
an effect.

This specific heat data is consistent with a spin glass in that the
observed feature is a broad maximum with no pronounced anomalies.
However, it is not clear how the unusual spin liquid or ``anti-glass'' state observed
with ac susceptibility~\cite{Reich1990,Ghosh2003} should manifest
itself in specific heat measurements.  Based on the numerical
simulations of Snider \emph{et al.}~\cite{Snider2005}, the measured
increase in entropy may indicate that the system is no longer a spin glass below
$x\simeq 0.08$ and is instead a spin liquid with many accessible
nearly-degenerate ground states.

\begin{acknowledgments}
We have benefited greatly from discussions with M.~J.~P. Gingras and
G.~M.~Luke. Also thanks to N.~F.~Heinig, S.~Meng, L.~Lettress and
N.~Persaud.  Thanks to J.~F.~Britten for a part of the X-ray
characterization.  Funding for this research was provided by NSERC,
CFI, MMO and Research Corporation grants.
\end{acknowledgments}

\end{document}